\documentstyle[12pt]{article}
\newcommand{\beq}{\begin{equation}}
\newcommand{\eeq}{\end{equation}}

\def\kap{{\textstyle{1\over{\kappa^2}}}}
\def\half{{\textstyle{1\over2}}}
\def\thre{{\textstyle{1\over3}}}
\def\twtr{{\textstyle{2\over3}}}

\def\quart{{\textstyle{1\over4}}}
\def\eigt{{\textstyle{1\over{2^3}}}}
\def\ninsix{{\textstyle{{{1}\over{3\cdot 2^6}}}}}

\def\p1half{{\textstyle{{{p+1}\over{2}}}}}

\def\23phalf{{\textstyle{{{23-p}\over{2}}}}}

\begin{document}
\thispagestyle{empty}
\begin{titlepage}

\bigskip
\hskip 3.7in{\vbox{\baselineskip12pt
\hbox{PSU-TH-248}\hbox{hep-th/0201129}}}

\bigskip\bigskip\bigskip\bigskip
\centerline{\large\bf
Nonperturbative Type I--I$^{\prime}$ String Theory}

\bigskip\bigskip
\bigskip\bigskip
\centerline{\bf Shyamoli Chaudhuri\footnote{On leave of absence from Penn State, 
Sep--Dec 2001. Part-time Visitor, YITP \& Kyoto Univ,
Fall 2001.}
}

\medskip
\medskip
\medskip

\centerline{Physics Department}
\centerline{Penn State University}
\centerline{University Park, PA 16802, USA}

\bigskip
\medskip
\centerline{Yukawa Institute for Theoretical Physics}
\centerline{Department of Physics}
\centerline{Kyoto University}
\centerline{Kyoto 606-8502, Japan}

\date{\today}

\bigskip\bigskip
\begin{abstract}
We propose a nonperturbative framework for the $O(32)$ type I open
and closed string theory. The short distance degrees of freedom are bosonic
and fermionic hermitian matrices belonging respectively to the adjoint and 
fundamental representations of the special unitary group $SU(N)$. We identify 
a closed matrix algebra 
at finite $N$ which corresponds to the Lorentz, gauge, and supersymmetry algebras 
of the large $N$ continuum limit. The planar reduction of our matrix theory coincides 
with the low energy spacetime effective action of the $d$$=$$10$ type I $O(32)$
unoriented open and closed string theory.  We show that matrix $T$-duality transformations
can yield a nonperturbative framework for the $T$-dual type I$^{\prime}$ closed string
theory with 32 D8branes. We show further that under a strong-weak coupling duality
transformation the large $N$ reduced action coincides with the low energy spacetime
effective action of the $d$$=$$10$ heterotic string, an equivalence at leading order
in the inverse string tension and with either gauge group $Spin(32)/{\rm Z}_2$ or
$E_8$$\times$$E_8$. Our matrix formalism has the potential of providing a 
nonperturbative framework encapsulating all of the weak coupling limits of M theory.
\end{abstract}

\end{titlepage}

\vskip 0.1in
The consistent perturbative quantization of Yang-Mills theories with chiral matter
fields coupled to gravity provided by either the type I or heterotic string
theories--- at weak coupling, and in a framework which is anomaly-free, and both
ultraviolet and infrared
finite, is a remarkable achievement \cite{gs,ghmr}. Nonetheless, in the
absence of a nonperturbative framework, there remains open both the question of 
fundamental principle \cite{polbook}, and the danger that the
properties of the nonperturbative ground state of string theory turn out to 
be drastically different
at even a qualitative level \cite{wit,dbrane}. In addition, many outstanding
problems of particle physics, such as vacuum selection, spacetime supersymmetry
breaking, and the generation of hierarchies in the Standard Model, are widely assumed
to only find resolution in nonperturbative physics at very high energy scales,
possibly even Planckian. Thus, it is a matter of crucial importance to discover the
correct framework for nonperturbative string theory.

\vskip 0.1in
Candidates for such a nonperturbative framework have been proposed based on the
large $N$ quantum mechanics of D0branes \cite{dbrane,bfss}: \lq\lq particles"
carrying both electric, and solitonic, charge, that have been proposed as the
fundamental constituents of both the solitons and strings of M theory in \cite{bfss}. 
An alternative starting point is the planar reduction of ten-dimensional super 
Yang-Mills theory giving the zero-dimensional matrix action of \cite{ikkt}. A third, 
and closely related, proposal is matrix string theory, based on the large $N$ limit 
of two-dimensional $SU(N)$ gauge theory \cite{dvv}. All three proposals are Wilsonian 
cut-off theories \cite{wilson}. The implicit
assumption has been made that off-diagonal $(1/N)$ quantum corrections to the 
planar action representing massive terms will, upon integrating out, provide missing 
relevant interactions in the classical action required in order to match correctly 
with the known infrared limit \cite{wit,polbook}. 
While the detailed matching between supergravity and matrix computations is difficult, 
with considerable room for ambiguity, the success of either proposal in meeting this
crucial consistency check remains in doubt \cite{dine}. 

\vskip 0.1in
More seriously, a clear-cut 
demonstration of spacetime Lorentz invariance in the large $N$ limit is lacking in 
both the BFSS model and in matrix string theory as a consequence of an inherently 
light-cone formulation which obscures many points of physics \cite{bfss,dvv}. The IKKT
model has different short-comings. An explicit demonstration of dynamically selected 
eigenvalue configurations characterizing the large $N$ ground state has given evidence 
for a nonperturbative mechanism for vacuum selection with fewer than 
ten noncompact spacetime dimensions \cite{nishimura}. While incompletely understand at 
the moment, this is intriguing. On the other hand, the arguments for making contact 
with the $\cal N$$=$$2$ spacetime supersymmetries of the ten-dimensional IIB string, 
and with its Dstring ground states, are shaky. A clear separation of Ramond-Ramond 
sector solitons from the Neveu-Schwarz sector solitons of the supersymmetric gauge 
theory has not been made. Perhaps the most significant aspect of the IKKT proposal, we
will argue, is the insight it offers into the dilemna of how to preserve manifest 
spacetime diffeomorphism invariance in a fully quantum theory of gravity.

\vskip 0.1in
The IKKT action is the dimensional reduction of $d$$=$$10$ ${\cal N}$$=$$1$ 
$SU(N)$ supersymmetric 
Yang-Mills theory to a point \cite{ek,ikkt}, or \lq\lq event", in spacetime. Not
surprisingly, it coincides with the classical worldvolume action of $N$ coincident 
Dinstantons \cite{dbrane,polbook}. While this is hardly the motivation for the IKKT 
proposal, we comment that it provides a simple understanding for the absence of a 
mass scale in the IKKT matrix action \cite{ikkt}: the Dpbrane tension scales as 
$\tau_p $ $=$ $O ( g^{-1} \alpha^{\prime -(p+1)/2} )$, dimensionless when $p$$=$$-1$ 
\cite{dbrane}. It also provides an understanding of extensions to the IKKT action
since such terms should have a natural interpretation in terms of the full quantum 
non-abelian Born-Enfield action \cite{polbook}. We will, however, take 
a somewhat different path 
towards a candidate nonperturbative matrix theory by first incorporating a crucial feature 
absent in both the BFSS and IKKT matrix 
actions. Namely, we replace the $N$$\times$$N$ fermionic, Grassmann-valued, matrices 
with Grassmann-valued matrices transforming in the $N$ and $\bar N$ fundamental 
representations of the $SU(N)$ group. This allows us to incorporate chirality into a 
matrix theory with sixteen supercharges, giving rise to sixteen component spinor 
fields in the large $N$ limit that can simultaneously satisfy both the Majorana and Weyl 
conditions of the $d$$=$$10$ $\cal N$$=$$1$ SYM-supergravity theory 
\cite{fs,gs,ghmr,br}.

\vskip 0.1in
Motivated by this insight, we consider a $SU(N)$ invariant hermitian matrix action 
with sixteen supercharges. The bosonic matrices live in the adjoint representation 
of the special unitary group, while the sixteen component chiral fermionic matrices 
transform in the $N$-dimensional fundamental representations of $SU(N)$. The components 
of a bosonic hermitian matrix belong to the field of real, or complex, numbers while 
those of a fermionic matrix belong to the field of anticommuting Grassmann 
numbers. The continuum large $N$ limit of our theory is defined as follows: we take 
$\epsilon$ $=$ $M_{11}^{-1}$ $\to$ $0$, where $M_{11}$ is the eleven-dimensional Planck 
mass, such that the length scale $\epsilon N$ is held fixed, providing a 
short distance cutoff in our theory \cite{wilson}. The continuum limit is therefore 
an expansion about a ground state with diagonal entries for all matrices,
such that in the large $N$ limit objects in the adjoint and 
fundamental representations evolve, respectively, 
into the continuum bose and fermi fields of a continuum action.
This continuum action will be matched with the spacetime low energy effective action 
of the heterotic-type I string theory at leading order in the $\alpha^{\prime}$ 
expansion. The short distance cutoff of the large $N$ theory can be 
expressed in terms of the ten-dimensional closed string tension by the relation: 
$\epsilon N $ $=$ $N e^{\twtr \Phi_0} \alpha^{\prime 1/2}$, where $\Phi_0$ is
the vev of the continuum dilaton field in the large $N$ ground state.

\vskip 0.1in
An important distinction from \cite{bfss,dvv} is incorporation of a key feature of 
the original
motivation behind the Eguchi-Kawai reduction of a large $N$ gauge theory \cite{ek}: the 
necessity for the explicit appearance of a spacetime continuum will be abandoned. Invoking 
translational invariance in the noncompact directions and the gauge symmetries will
enable reduction of the number of dynamical degrees of freedom in the nonperturbative 
supersymmetric gauge-gravity theory to those on a single site of the spacetime lattice. 
Notice that spacetime diffeomorphism invariance follows automatically in any 
nonperturbative matrix formulation of a quantum theory of gravity based on 
{\em zero-dimensional} matrices. It is only in the specification of the 
ground state, namely, the large $N$ limit about which we expand when computing the 
$O(1/N)$ quantum corrections arising in the matrix path integral, that diffeomorphism 
invariance is spontaneously broken.  Upon taking the large $N$ continuum limit, we must 
refer all quantum fields to a fixed background spacetime metric. Thus,
the perturbative string theories, and their low energy effective field theory 
limits, describe the long-distance fluctuations of matter-energy in a spacetime continuum, 
but with respect to a fixed background metric, and in a ground state with spontaneously 
broken diffeomorphism invariance. Notice that we preserve the spirit of 
Einstein's classical theory of general relativity: spacetime geometry is replaced by an 
equivalent matter-energy distribution together with the Einstein equations \cite{einstein}. 
But given both the matter fields and a specific solution to the classical
equations enables recovery, at least in principle, of a corresponding spacetime metric. 
This is also true for the nonperturbative matrix quantum theory of gravity.
The target spacetime geometry is abandoned in favour of matrix degrees of freedom.
However, in any large $N$ ground state of the theory, diffeomorphism invariance will
be {\em spontaneously} broken: a necessary feature of the long-distance effective description 
of the gauge-gravity interactions provided by the string theories \cite{gs,ghmr,polbook}. 

\vskip 0.1in
In this paper, we arrive at a proposed matrix action for nonperturbative string theory 
by the process of {\em inference}, from a knowledge of both the target-space, and 
strong-weak coupling, 
dualities \cite{wit,dbrane}, and the specific form of the low energy spacetime effective 
action of the type I open and closed unoriented string theory \cite{fs,ghmr,br,polbook}. 
Our methodology obviates the difficulties encountered in making a precise match between 
S-matrix amplitudes of the matrix model and of supergravity theory \cite{dine}. It 
simultaneously incorporates in the large $N$ continuum limit both the matter 
(Yang-Mills) and supergravity sectors of the anomaly-free ten-dimensional type I string.
It also gives a clear demonstration of Lorentz invariance and of manifest 
supersymmetric and non-abelian gauge covariance in the large $N$ continuum theory. But, 
best of all, we find that the classical matrix action thereby obtained has an elegant
and simple structure that can be motivated largely from first principles, and by symmetry 
considerations alone. Perhaps this is not surprising given the aesthetic use of symmetry 
principles--- both kinematic and dynamical, in the original formulation of perturbative 
string theory \cite{gs,ghmr,polbook,power}, now appearing in the large $N$ limit of a hermitian 
matrix action.

\vskip 0.15in
Our starting point is an action with $N$$\times$$N$ bosonic hermitian matrices
transforming in the adjoint representation of the special unitary group, coupled 
to sixteen component fermionic hermitian matrices in the $N$-dimensional 
fundamental representations of $SU(N)$. In what follows, we work in natural units 
setting $\hbar$ $=$ $c$ $=$ $1$. The $SU(N)$ matrix variables may carry, in addition, 
both Lorentz and nonabelian group indices. Matching the short-distance cutoff
$\epsilon N$ with the string scale, we infer the form of matrix action required 
in order that the large $N$ limit yields a continuum effective field theory action 
exhibiting ${\cal N}$$=$$1$ spacetime supersymmetry, local Lorentz invariance, 
and Yang-Mills gauge symmetry, and with the anomaly-free massless field content of
the $O(32)$ type I string \cite{wilson,polbook}. For covenience, we denote the 
finite-dimensional Yang-Mills group as the generic group $\bf G$, of rank $r_G$, 
and dimension $d_G$, although the reader can assume for the purposes of this paper
that $\bf G$ is the rank 16 group $SO(32)$ (or $E_8$$\times$$E_8$, see \cite{power}). 
As is well-known, upto suitable identifications under heterotic-type I strong-weak 
coupling duality \cite{wit,pw,polbook}, the large $N$ continuum action 
also corresponds to the low energy spacetime effective action of the $d$$=$$10$
$Spin(32)/{\rm Z}_2$ heterotic string \cite{ghmr,polbook}.

\vskip 0.1in
The inner product in the Hilbert space for any hermitian matrix variables is defined by
the direct multiplication of matrices, while taking care to preserve $SU(N)$ invariance. 
Since the matrix variables can, in addition, carry Lorentz indices, a matrix 
ordering prescription compatible with the continuum Lorentz transformations of the 
fields obtained in the large $N$ limit is necessary. Specifically, we must distinguish left-
and right- matrix multiplication within any given bilinear of matrices. We fix the 
ordering ambiguity by recognizing the $SL(2,C)$ decomposition of each Lorentz tensor: an
object in the $(n,0)$, or $(0,{\bar {n}})$, representation is a left-multiplier, while
the $(0,n)$, or $({\bar{n}} , 0)$ representation multiplies from the right. Thus, anti-spinors 
preceed spinors in any matrix bilinear, and contravariant tensors preceed covariant 
tensors. By \lq\lq spinor" and \lq\lq anti-spinor" here, we mean fermionic, Grassmann-valued, 
matrices: objects transforming in the fundamental and anti-fundamental, the $N$ and 
${\bar{N}}$ representations of $SU(N)$, evolving, respectively, into spinor and anti-spinor 
continuum fields in the large $N$ limit. Likewise, for the generic covariant and 
contravariant matrix Lorentz \lq\lq tensors".
It is helpful to write down explicit expressions for the simplest cases. Consider the 
inner product for a matrix Lorentz \lq\lq tensors" defined so as to exhibit manifest 
invariance under finite $N$ matrix Lorentz transformations. We use tangent space indices 
whenever possible, avoiding explicit appearance of the matrix \lq\lq vierbein", $E^{\mu}_a$, 
both in the action, and in the inner product. Tangent space indices are raised and lowered 
by the $\eta^{ab}$ and $\epsilon^{ab}$ symbols. We note the following simple identities 
for generic covariant and contravariant matrix \lq\lq tensors":
\begin{eqnarray}
\eta^{ab} =&& E^{a \mu} E^{b}_{\mu},  \quad  \eta_{ab} = E^{\mu}_a E_{b \mu} , \quad
{A}^{\mu} A_{\mu} = {A}^a E^{\mu}_a E_{b \mu} A^b = {A}^a A_a \cr
{A}^{\mu} =&& A^a E^{\mu}_a , \quad ~ A_{\mu} = E_{b \mu} A^b , \quad
~ {\bar{\psi}}_{\mu} ={\bar{\psi}}_{a}  E^{a}_{\mu} , \quad
~ \psi_{\mu} = E^a_{\mu} \psi_{a} \cr
{A}^{\mu\nu} A_{\mu\nu} =&& {A}^{ca} E^{\nu}_a E^{\mu}_c E_{d \mu} E_{b \nu} A^{db} =
{A}^{ca} E^{\nu}_a \eta_{cd} E_{b \nu} A^{db} = {A}^{ca} \eta_{ab} {A_c}^b  =
{A}^{ca} A_{ca}
\quad ,
\label{eq:raise}
\end{eqnarray}
and likewise for the higher rank cases. The ordering of matrices in the
inner product is defined unambiguously. Notice the natural correspondence
between the trace of a given matrix bilinear of $SU(N)$ matrices with the Lorentz 
invariant inner product for a continuum field obtained in the large $N$ limit 
\cite{bilocal,ikkt}. We have:
\begin{eqnarray}
({\bar{\Psi}} , \Psi^{\prime}) ~\equiv&& {\rm tr}~ {\bar{\Psi}} \Psi^{\prime} \quad
\quad ~ \to~
\int d^{10} x ~ {\bar{\Psi}}(x) \Psi^{\prime}(x) \cr
( A_{[1]} , A^{\prime}_{[1]} ) ~\equiv&&
{\rm tr} ~ {A}^{\mu} A^{\prime}_{\mu} \quad
~~ = ~ {\rm tr} ~ {A}^{a} A^{\prime}_{a} \quad
\quad \to~
\int d^{10} x  ~ A^{a} (x)  A^{\prime}_{a} (x) \cr
( A_{[2]} ,  A^{\prime}_{[2]}) ~\equiv&&
{\rm tr} ~ {A}^{\mu\nu } A^{\prime}_{\mu\nu} \quad
= ~ {\rm tr} ~ {A}^{ab } A^{\prime}_{ab} \quad
~\to~ \int d^{10} x ~ A^{ab} (x)  A^{\prime}_{ab} (x)
\quad ,
\label{eq:fermip}
\end{eqnarray}
and likewise for higher rank cases. The trace denotes a sum over the diagonal 
elements of the composite matrix operator. The continuum fields $\Psi(x)$, 
$A_{a}(x)$, and $A_{ab }(x)$, transform, respectively, as a d=10 tangent space 
spinor, vector, and two-form tensor field. Notice that each matrix \lq\lq tensor" 
appearing in a matrix bilinear is rescaled by a power of the matrix \lq\lq 
vierbein", $E^{1/4}$, consistent with the appearance of diffeomorphism invariance 
in the large $N$ continuum limit.

\vskip 0.1in
\noindent To illustrate the basic methodology by which we will proceed, recall 
the well-known case of the $N$$\times$$N$ hermitian one-matrix model \cite{bipz}:
\begin{equation}
{\cal S_B} = {\rm Tr} ~ \Phi \Omega^a \Omega_a \Phi   , 
\quad \Delta \equiv \Omega^a \Omega_a ,
\quad \Omega_a \Phi_n = \lambda^{1/2}_{na} \Phi_n  , 
\quad n = 1,\cdots , N \quad . 
\label{eq:scalar}
\end{equation} 
We are assuming here that $\Delta$$=$$\Omega^a \Omega_a$ is a 
self-adjoint linear operator $\Delta$, with orthonormalized eigenfunctions, 
$\{ \Phi_n \}$, and eigenvalues, $\{ \lambda_n \}$. This will indeed be true for
all of the \lq\lq kinetic" terms in our matrix action. In the large $N$ limit, we 
will make the usual extension to a Hilbert space with continuous eigenfunctions 
$\{ \phi(\lambda ; {\bf x}) \}$. Note that the dependence on the underlying 
target space, ${\bf x}$, is only implicit; we work in a generalized momentum 
basis, or with suitable phase space variables. The transition from 
finite $N$ to infinite $N$ could be ill-defined. Given 
the nature of the eigenvalue spectrum of $\Delta$ in the large $N$ limit, 
appropriate low and high momentum regulators may be needed in order to make the 
integral well-defined. Alternatively, one can imagine modifying the 
large $N$ limit, as was done in the well-known case of the double-scaling limit 
for string theories with superconformal matter of central charge 
$c$$\le$$3/2$ \cite{mat}. In the cases $c$$<$$3/2$, where there is no 
one-dimensional embedding superspace, the eigenvalue density variable obtained
in the large $N$ limit can be identified with the super-Liouville modes 
expected from two-dimensional superconformal supergravity \cite{poly,mat}. This 
equivalence followed from the remarkable isomorphism observed between the graphical 
expansion for triangulated random surfaces and the dual planar graph expansion of 
the one-matrix model with generic polynomial potential \cite{bipz,mat}. Notice that 
no symmetry other than $SU(N)$ invariance, employed here in order to diagonalize 
the matrix and thereby decouple the angle variables \cite{bipz}, has been invoked 
in analytically treating the one-matrix path integral. What happens if the matrix 
action is characterized by additional symmetry? 

\vskip 0.1in
In the case of gauge theories, recall that we can invoke gauge invariance 
in performing the path integration over orbits of the gauge group, extracting the 
group volume a la Faddeev-Popov, and accounting separately for any potential infrared 
divergences arising from the integration over the gauge slice.
In the case of the hermitian one-matrix model \cite{bipz}, this was
straightforwardly implemented both at finite and at infinite $N$, since both the 
action and the inner product localizes on the diagonal elements of the $N$$\times$$N$ 
matrix. In the presence of interactions coupling two, or more, species of matrices, 
an analytic treatment of zero-dimensional matrix path integrals is less obvious, 
although the conceptual underpinnings are similar \cite{bipz}.  
More generally, in order to perform the path integration over matrix variables 
we must specify an inner product invariant under the extended symmetry group 
${\cal G}$ of matrix transformations that leave the classical action invariant: 
let ${\cal G}$ define the finite $N$
remnant of the Lorentz, gauge, and supersymmetry invariances of the
continuum large $N$ action. Such finite $N$ symmetry transformations have 
nontrivial overlap with the group of $SU(N)$ rotations. Since they constitute
 a \lq\lq gauge invariance" of our matrix theory, ${\cal G}$ invariance
should be manifest both in the classical action, and in the inner
product defined on the Hilbert space of matrix \lq\lq tensors".
We must be careful in defining the group invariant measure in
the matrix path integral, where by \lq\lq group" we mean here the
semi-direct product
group $\cal G$$\times$$SU(N)$ acting on the hermitian matrix \lq\lq tensors".
Since every term in the continuum large $N$ action
is required to be singlet under both matrix Lorentz, $\cal L$, and matrix
Yang-Mills, $\bf G$, transformations, we will identify matrix composites
invariant under the finite $N$ matrix manifestation of these symmetries.
We comment that this procedure is reminiscent of equivariant
localization \cite{niemi}, an observation of likely note considering the
matrix theory can in many respects be considered topological, in the sense 
that it has no intrinsic length scale. Having briefly clarified the 
motivation for identifying matrix algebra analogs of the continuum Lorentz, 
gauge, and supersymmetry invariances of the effective field theories 
describing the large $N$ ground state, we proceed to an analysis of the
classical matrix action.

\vskip 0.1in
Our expression for the classical matrix action will be manifestly invariant
under ${\cal L}$$\times$$\bf G$. With guidance from the low energy continuum 
type I string effective action \cite{fs,br,polbook}, it is not difficult to 
infer its form:
\begin{eqnarray}
{\cal S} &&=~
\kap \left ( {\bar \psi}_{a} \Gamma^{abc} D_{b} \psi_{c}
~-~ 4 {\bar{\lambda}} \Gamma^{ab} D_{a} \psi_{b}
~-~ 4 {\bar{\lambda}} \Gamma^{a} D_{a} \lambda \right )
~+~ g^2 e^{\Phi}~ {\bar \chi}^i \Gamma^{a} D_{a} \chi^i
\cr
\quad\quad  &&\quad\quad
~+~ g^2 e^{\Phi }~ {F}^{ab} F_{ab}
~+~ \kap \left ( {\cal R}
~-~ 4~ {\Omega}^{a} \Phi ~ \Omega_{a} \Phi
~+~ 3 e^{2 \Phi} ~ {H}^{abc} {H}_{abc} \right )
\cr
\quad && \quad\quad\quad\quad
~+~   {\cal S}_{\rm 2-fermi}  ~+~ {\cal S}_{\rm 4-fermi}  \quad .
\label{eq:hmat}
\end{eqnarray}
Spinor and $SU(N)$ indices have been suppressed in this expression,
and the notation is as follows. In comparing with the low energy
string effective action \cite{br}, note that we have absorbed the
overall factor of $E^{1/2} e^{- 2 \Phi }$ in the definition
of the $SU(N)$ trace, also omitting an overall minus sign in the action.
This will be of importance when we give a detailed prescription for
taking the continuum limit with some, or all, coordinates noncompact.
It is conventional in the literature to write the continuum low energy
spacetime effective action with an additional overall factor of $\half$
\cite{polbook}, also dropped in our matrix action.
Notice that explicit dilaton dependence in the measure is restricted to
the kinetic terms for the Yang-Mills, and two-form, matrix \lq\lq tensors".
This is a result of our rescaling of the $SU(N)$ trace.

\vskip 0.1in
In the expression above, $\chi^{i\alpha} $, ${\bar{\chi}}^{i\alpha}$,
denote Grassmann-valued fermionic matrices in the $N$, ${\bar{N}}$,
representations of $SU(N)$. The indices, $i$$=$$1$, $\cdots$, $r_G$,
simultaneously labels a fundamental representation of the Yang-Mills group 
$\bf G$, while $\alpha$$=$$1$, $\cdots$, $16$, labels sixteen distinct
Grassmann-valued fermionic matrices, evolving in the large $N$ continuum 
limit into the sixteen components of a Majorana-Weyl spinor field.
Likewise, $\psi^{\alpha}_{\mu }$, denotes Grassmann-valued fermionic
matrices evolving in the large $N$ limit into the components of a 
Lorentz spinor-vector field in ten dimensions.
Finally, we have the matrix representatives of the dilatino field,
also in Grassmann-valued $SU(N)$ fundamental representations, 
$\lambda^{\alpha}$. In the continuum limit, $\chi^i$, 
$\psi_{\mu}$, and $\lambda$, yield, respectively, the
gaugino, gravitino, and dilatino fields of the d=10
${\cal N}$$=$$1$ SYM supergravity action. The $SU(N)$ matrices 
$F_{ab}$, $H_{abc}$, $\cal R$, and $\Phi$ are, respectively,
finite $N$ matrix representatives of
the Yang-Mills tensor, the shifted antisymmetric
three-form field strength corresponding to the
two-form potential $C_{[2]}$, plus Chern-Simons term for the
Yang-Mills potential, $A_{[1]}$, the Ricci
curvature, and the dilaton scalar continuum fields.

\vskip 0.1in
The matrix operator $D_{a} \psi_{b}$ is required to
evolve into a rank two covariant tensor in the continuum large
$N$ limit. In general, $\Gamma^a D_a$ may be expressed as an
expansion in the complete set of independent Lorentz structures
in the type II theories with sixteen component spinors:
\begin{equation}
(\Gamma^a D_{a})_{\alpha\beta} \equiv \Omega_a (\Gamma^a)_{\alpha\beta}
+ \Omega_{ab} ( \Gamma^a \Gamma^{b})_{\alpha\beta}
+ \Omega_{abc} (\Gamma^a \Gamma^{bc})_{\alpha\beta}
+ \Omega_{abcd} (\Gamma^a \Gamma^{bcd})_{\alpha\beta}
+ \Omega_{abcde} (\Gamma^a \Gamma^{bcde})_{\alpha\beta}
+ ~ {\rm duals} \quad .
\label{eq:covmateq}
\end{equation}
In the large $N$ continuum limit, the matrix operator $\Gamma^a D_a$
is required to evolve into the appropriate bosonic, or fermionic,
super-covariant derivative, with appropriate couplings, in general, to
both spin connection and nonabelian vector potential. In principle,
it could couple as well to the two, or magnetic dual six-form, gauge
potentials of the nonperturbative type I
string theory \cite{dbrane,polbook}. Note that the dual couplings are not,
however, necessitated by closure of the matrix Lorentz algebra. 
Most generally, including only the full spectrum of even potentials in the 
IIB theory, we could write:
\begin{eqnarray}
\Gamma^a D_{a}
\to&&
\left  ( ( {\bf 1} + C_{(0)}(x) ) \partial_{a}  +
 A_{a}^{j} (x) \tau^{j} \right ) \Gamma^a
+ \left ( A_{ab} (x) + C_{ab} (x) \right ) \Gamma^a \Gamma^{b}
+  \omega_{abc} (x) \Gamma^a \Gamma^{bc}
\cr
\quad &&\quad
+  C_{abcd} (x) \Gamma^a \Gamma^{bcd}
+  \omega_{abcde} (x) \Gamma^a \Gamma^{bcde}
\quad .
\label{eq:covmat}
\end{eqnarray}
The one-form, $A_{a}$ $\equiv$ $A_{a}^j \tau^{j}$, is the familiar Yang-Mills
gauge potential. The $\tau^{j}$ are hermitian generators of the Yang-Mills
gauge group $\bf G$, the $f_{ijk}$ are its structure constants,
$[ \tau^i , \tau^j ] $$=$$i f^{ijk} \tau^k$, and
$g$$=$$e^{\Phi}$ is the dimensionless open string coupling. We
note that this prescription for
incorporating nonabelian gauge symmetry is similar in spirit to the suggestion
in \cite{ikkt} that an $SU(K)$ Yang-Mills symmetry arises whenever eigenvalue
configurations decompose naturally into block-diagonal clusters of
$K$ eigenvalues. Finally, $\omega_{abc}[e(x)]$ is
the ordinary spin-connection including the dilaton dependent piece,
expressed in terms of the vierbein in the first-order formalism for
the low energy type I string effective action \cite{fs,polbook}.

\vskip 0.1in
The tensors, $C_{ab}$, and $C_{abcdef}$, represent gauge
potentials coupling to, respectively, D1branes, or
D5branes, when present in the IB large $N$ ground state \cite{dbrane,polbook}. 
The $\Phi$ dependence in their kinetic terms in the effective action will 
distinguish them from antisymmetric gauge potentials with the same number 
of Lorentz indices \cite{polbook}. The two-form matrix \lq\lq tensor", 
$C_{[2]}$, although not required, is consistent with both the matrix 
supersymmetry and matrix Lorentz algebra as will be shown below.
Finally, recall that the fluctuating parts of the continuum fields,
$A_{ab}$, $C_{abcd}$, and $C_{(0)}$, are absent in the ten-dimensional
type I unoriented string. We should emphasize that none of the 
antisymmetric tensor potentials are {\em necessary} for closure of the 
Lorentz algebra. Instead, they represent additional vacuum charges
consistent with closure of the Lorentz algebra. These \lq\lq Ramond-Ramond" 
charges \cite{dbrane}--- originally named for the fermionic vacuum in the 
sector of the perturbative string theory with periodic, or Ramond, boundary 
conditions on both left- and right-handed worldsheet fermions \cite{polbook}, 
become necessary only upon application of the weak-strong coupling 
duality principle to the string vacuum state \cite{wit,dbrane,polbook}.

\vskip 0.1in
Notice that the five-index antisymmetric matrix \lq\lq tensor" is new, 
unanticipated from field theory considerations, although clearly permitted 
by symmetry considerations alone. The necessity of including such a 
term in the classical matrix action comes from closure of the matrix 
Lorentz algebra when acting on sixteen-component Majorana-Weyl spinors: 
a five-index matrix commutator, $[\Omega_{abc} ,
{\bf L}_{de}]$, appears naturally upon Lorentz transformation of the
spin-connection.  Thus, closure of the matrix Lorentz algebra for a
generic Majorana-Weyl spinor,
$\Psi$, {\em requires inclusion of a five-index antisymmetric matrix},
$\omega_{abcde}$, in addition to the usual spin-connection. The
supercovariant derivative operator may be defined as follows:
\begin{equation}
\Gamma^a D_a \Psi = \Gamma^a \left ( \Omega_a + \half \Omega_{abc} \Gamma^{bc}
+ \quart \Omega_{abcde} \Gamma^{bcde} \right ) \Psi  \quad .
\label{eq:covfive}
\end{equation}
It is easy to verify that in the presence of a matrix two-form 
\lq\lq tensor" potential, $C_{ab}$, the nontrivial commutator with 
${\bf L}_{de}$ implies a necessity for both non-vanishing $C_{[0]}$,
and $C_{[4]}$, matrix \lq\lq tensors". In the continuum large $N$ 
limit, the matrix \lq\lq field strengths" assume the familiar form 
of the corresponding fields:
\begin{eqnarray}
R_{\mu\nu}^{ab} [\omega_{ab}^{\mu} ] ~\to&&
\partial_{\mu} \omega_{\nu}^{ab}(x) -
\partial_{\nu} \omega_{\mu}^{ab}(x) +
\omega_{\nu}^{ac} (x) \omega_{\mu c}^b (x) -
\omega_{\mu}^{ac} (x) \omega_{\nu c}^b (x) \cr
F^i_{\mu\nu} [A_{1}] ~\to&& \partial_{\mu} A^i_{\nu} (x)
- \partial_{\nu} A^i_{\mu} (x) +  f^{ijk} A^j_{\mu}(x) A^k_{\nu} (x) \cr
H_{\mu\nu\rho}[C_{\mu\nu},A_{\mu}] ~\to&&
\partial_{[\mu} C_{\nu\rho]} -
  {\rm tr} ~ (A_{[\mu}\partial_{\nu} A_{\rho]}
  - {{2}\over{3}} A_{[\mu} A_{\nu} A_{\rho ]} )
\quad .
\label{eq:fields}
\end{eqnarray}
$C_{[2]}$ is the two-form gauge potential of the type I string. The antisymmetric 
two-form matrix potential, $A_{ab}$, which ordinarily couples in the large 
$N$ limit to fundamental, heterotic, or type II, oriented closed strings, 
is restricted to take constant values alone in ground states of the unoriented
type I string. Likewise, acting on scalar, vector, and two-form, potentials, 
$\Omega_a$ is required to evolve into the familiar nonabelian gauge covariant
derivative in the large $N$ continuum limit:
\begin{equation}
\Omega_a
\to  \partial_a + A^{j}_a \tau^{j} \quad .
\label{eq:gaugeciv}
\end{equation}
It is easy to verify closure of the matrix Lorentz algebra in the absence of
$p$-form gauge potentials, $p$$\ge$$2$, with this definition for $\Omega_a$.
Closure corresponds, in this case, to the full ten-dimensional Poincare group
of spacetime symmetries characterizing the type I ground state with ten-dimensional
${\cal N}$$=$$1$ supergravity and $O(32)$ Yang-Mills fields in the large $N$ 
continuum limit.

\vskip 0.1in
Returning to a complete description of the classical matrix action, we must
include crucial two-fermion and four-fermion terms required by supersymmetry 
\cite{fs,br}. With guidance from the continuum low energy type I string effective 
action \cite{br}, and keeping in mind the overall minus sign and factor of 
$\half$ described above, we infer that the 2-fermi terms take the form:
\begin{eqnarray}
 {\cal S}_{\rm 2-fermi} &&=
2~ {\bar{\psi}}_{a} \Gamma^{a} \psi_{b} ( \Omega^{b} \Phi )
~-~ 4 ~{\bar {\psi}}_{a} \Gamma^{b} \Gamma^{a} \lambda ( \Omega_{b} \Phi)
\cr
&&\quad\quad - \eigt H^{def} \left [
{\bar \psi}_{a} \Gamma^{[a} \Gamma_{def }\Gamma^{b]} \psi_{b}
~+~ 4 ~{\bar \psi}_{a} \Gamma^{a}_{def } \lambda
~-~ 4 ~ {\bar\lambda} \Gamma_{def } \lambda
~+~ g^2 e^{\Phi} {\bar \chi} \Gamma_{def} \chi \right ]  \cr
&&\quad\quad\quad
~+~ \quart g^2 e^{\Phi} ~ {\bar \chi}^i \Gamma^{d} \Gamma^{ab}
(\psi_{d} +  \thre \Gamma_{d} \lambda ) F_{ab}^i
\quad .
\label{eq:2fermat}
\end{eqnarray}
Likewise, the 4-fermi terms in the matrix classical action take the form:
\begin{eqnarray}
{\cal S}_{\rm 4-fermi} &&=
\ninsix {\bar\psi}^{f} \Gamma^{abc} \psi_{f}
 \left (  {\bar\psi}_{d} \Gamma^{d} \Gamma_{abc} \Gamma^{e} \psi_{e}
      + 2 ~ {\bar{\psi}}^{d} \Gamma_{abc} \psi_{d}
        - 4 ~ {\bar{\lambda}} \Gamma_{abc} \lambda
          - 4~ {\bar{\lambda}} \Gamma_{abc} \Gamma^{d}\psi_{d}
                   \right )
\cr
&&\quad\quad
~+~ \ninsix g^2 e^{\Phi} ~ ( {\bar{\chi}} \Gamma^{abc} \chi ) \left (
{\bar{\psi}}_{d} ( 4~ \Gamma_{abc} \Gamma^{d} + 3 ~ \Gamma^{d} \Gamma_{abc} ) \lambda
   - 2~ {\bar{\lambda}}\Gamma_{abc}\lambda - 3\cdot 2^3 ~ H_{abc}  \right ) .
\label{eq:4fermimat}
\end{eqnarray}
As before, it can be verified that matrix Lorentz invariance holds for
both the 2-fermi and 4-fermi terms in the classical matrix action. The 
expression for ${\cal S}$ can be simplified and written in a more compact 
form by introducing $SU(N)$ vectors, $\Psi$, ${\bar{\Psi}} $, and the matrix 
operator, ${\cal D}$. The $(1$$+$$10$$+$${\rm d}_{\rm G})N$-component $SU(N)$
vectors transform as, respectively, 16-component right- and left-handed 
Majorana-Weyl \lq\lq spinors" under the matrix Lorentz group, and can be 
denoted as follows:
\begin{equation}
{\bar{\Psi}} \equiv ({\bar{\lambda}} , {\bar{\psi}}_{a} , {\bar{\chi}}^i )  ,
\quad \quad
\Psi \equiv (\lambda , \psi_{b} , \chi^j ) \quad .
\label{eq:fermions}
\end{equation}
Likewise, we assemble the independent Lorentz structures in the kinetic and 
two-fermi terms of ${\cal S}$ inside a matrix array of size, 
$(11$$+$${\rm d}_{\rm G})N $$\times $$ (11$$+$${\rm d}_{\rm G})N $.
%Explicitly, ${\cal D}$ is found to take the form:
%\begin{eqnarray}
%\quad {\cal{D}} =
% \left ( \begin{array} {c}
%   4~ \Gamma^{d} D_{d} - \half H^{def} \Gamma_{def}
%~~~~~~~~~~~~~ 4 ~ \Gamma^{db } D_{d}~~~~~~~~~~~~~~~~~~~~~~~~~0~~ \\
%\\
%   4~ \Gamma^{d} \Gamma^{a} \Omega_{d} \Phi
%~~~~~~~~~~~~- \Gamma^{adb} D_{d} -  2 ~\Gamma^{a} \Omega^{b} {\Phi}
%~~~~~~~~~~~~~~~~~~~~0~~\\
%~~~~+ \half~ H^{def} \Gamma^{a}_{def} ~~~~~~~~~~~~~
%    + \eigt ~ \Gamma^{[a} H^{def} \Gamma_{def}\Gamma^{ b ]}
%          ~~~~~~\\
%\\
%  - \quart g^2 ~ \Gamma^{d} \Gamma^{ab} \Gamma_{d} F_{ab}^i ~~~~~~~~~~~~~
%  - \quart g^2 ~ \Gamma^{d} \Gamma^{ab} F_{ab}^i ~~~~~~~~~~~
%              ~~~~~~~~~~ g^2 ( \Gamma^{d} D_{d}  + \eigt \Gamma^{def} H_{def} ) ~~
%       \end{array} \right ) .
%\nonumber
%\label{eq:dkinetic}
%\end{eqnarray}
%Note that spinor indices have been suppressed in these expressions, all others
%are explicit. 
Likewise, the four-fermi terms can be expressed in compact form
by introducing matrices, $\cal U$, $\cal V$, of size
$(11+{\rm d_G})N$$\times$$(11+{\rm d_G})N$, by referring to 
Eq.\ (\ref{eq:4fermimat}). Finally, we can define the matrix scalar 
\lq\lq Laplacian", $\Delta$ $\equiv$ $\Omega^a \Omega_{a}$. The matrix 
classical action can now be written in remarkably compact form:
\begin{equation}
{\cal S} ~=~ \half {\bar\Psi} {\cal D} \Psi +
  \quart ({\bar{\Psi}}{\cal U} \Psi )( {\bar{\Psi}} {\cal V} \Psi)
  + \half g^2 e^{\Phi} ~ F^{ab} F_{ab}
  + \half \kap ~ ( {\cal R}
     - \half \Phi \Delta  \Phi
     + 3 e^{2\Phi} ~ H^{abc} H_{abc} )
\quad .
\label{eq:compact}
\end{equation}
Under the matrix Lorentz group, $\Psi$ and $(\Psi)^{*t}$ transform,
respectively, in the $(0, \half )$ and $(\half , 0)$ representations.

\vskip 0.1in
The precise form of the matrix action given above has as large $N$ 
ground state the unoriented type I open and closed string theory in 
noncompact flat ten-dimensional spacetime \cite{fs,polbook}. As explained 
earlier, in order for the large $N$ ground state of the matrix theory to
correspond to a string vacuum carrying additional charges due to the
presence of Dbranes as part of the target spacetime geometry, we must 
include the appropriate kinetic and Chern-Simons terms for the matrix 
$p$-form potentials of the type I theory. We allow for the 
possibility of constant $A_{[2]}$. And, in addition to $C_{[2]}$,
incorporated in the shifted three-form field strength, we must include 
kinetic terms for all $C_{a_1 \cdots a_p}$, with $p$ even, and 
$0$ $\le$ $p$ $\le$ $10$ \cite{dbrane,polbook}. This is as required 
by the application of weak-strong coupling duality transformations on 
a given large $N$ ground state. Notice that in the matrix $T_9$-dual
type I$^{\prime}$ ground state, the corresponding $T_9$-dual matrix
$p$$\pm$$1$-form potentials will appear. Thus, the matrix action for
the $T_9$-dual type I$^{\prime}$ nonperturbative string is easily inferred
from a knowledge of the $T_9$-duality transformations of the corresponding
field in the large $N$ limit effective field theory ground state 
\cite{polbook}.

\vskip 0.1in
Finally, let us verify closure of the group of matrix
transformations that are the finite $N$ manifestation of the
Lorentz symmetries characterizing the large $N$ continuum limit.
We introduce an infinitesimal hermitian matrix, ${\rm L}_{ab}$,
antisymmetric under the interchange of tangent space indices
$a$,$b$. Keeping terms upto linear in ${\rm L}_{ab}$, it is easy
to verify that the matrix action ${\cal S}$ is invariant under
matrix Lorentz transformations which we define as follows:
\begin{eqnarray}
\delta \chi^i =&&  \Gamma^{ab} {\rm L}_{ab} \chi^i
, \quad
\delta {\bar{\chi}}^i = -  {\bar{\chi}}^i \Gamma^{ab} {\rm L}_{ab}
\cr
\delta \psi_{a} =&&  \Gamma^{bd} {\rm L}_{bd} \psi_a
    + {\rm L}_a^c \psi_c
, \quad
\delta {\bar{\psi}}_c = - {\bar{\psi}}_c \Gamma^{ab} {\rm L}_{ab}
    - {\bar{\psi}}_a {\rm L}^a_c
\cr
\delta \lambda =&& \Gamma^{bd} {\rm L}_{bd} \lambda
, \quad
\delta {\bar{\lambda}} =
- {\bar{\lambda}} \Gamma^{ab} {\rm L}_{ab}
\cr
\delta (\Gamma^a D_a ) =&& [ \Gamma^{ab} {\rm L}_{ab} , \Gamma^c D_c ]
\cr
\delta (\Gamma^{ab} D_b ) =&& [ \Gamma^{de} {\rm L}_{de} , \Gamma^{ab} D_b ]
+ {\rm L}^a_d \Gamma^{db} D_b
\cr
\delta (\Gamma^{abc} D_b ) =&& [ \Gamma^{de} {\rm L}_{de} , \Gamma^{abc} D_b ]
- [ \Gamma^{abd} D_b , {\rm L}_d^c ]
\cr
%\quad \delta  \Omega_{a} =&& {\rm L}_a^c \Omega_c ,
\quad \delta ( \Omega_{a} \Phi ) =&& [ {\rm L}_a^c , \Omega_c ] \Phi - \Omega_c {\rm L}^c_a \Phi ,
%\quad \delta  \Omega^a \Phi = - [ \Omega^c , {\rm L}_c^a ] ,
\quad \delta ( \Phi \Omega^a ) = - \Phi [ \Omega^c , {\rm L}_c^a ] +
 \Phi {\rm L}^a_c \Omega^c
%\quad \delta \Phi = 0
\quad .
\label{eq:lorent}
\end{eqnarray}
Notice that these transformation rules coincide with the usual
Lorentz transformations for the fields appearing in the large $N$
continuum limit. The reader can easily verify that the inner
product defined in Eq.\ (\ref{eq:fermip}) is
invariant under matrix Lorentz transformations.

\vskip 0.1in
Likewise, consider closure under matrix Yang-Mills transformations.
These will be defined so as to coincide with the usual nonabelian
gauge transformation rules for the
matter and gauge fields obtained in the large $N$ limit.
Consider a $d_G$-plet of
infinitesimal real matrices, $\alpha^j$, each of which
takes diagonal $N$$\times$$N$ form. The matrix Yang-Mills
transformations will be defined as follows:
\begin{eqnarray}
\delta \chi =&&  i \tau^j \alpha^j  \chi
, \quad
\delta {\bar{\chi}} = - i \tau^j {\bar{\chi}} \alpha^j
, \quad
\delta (g A_a^j \tau^j ) = [ \Omega_a , \tau^j \alpha^j ]
\cr
\delta ( \Omega_{a} \Phi ) =&& i \tau^j \alpha^j \Omega_a \Phi
, \quad \quad \delta ( \Phi {\hat{\Omega}}_{a} ) = - i \tau^j \Phi {\hat{\Omega}}_a
 \alpha^j
\quad .
\label{eq:gaugena}
\end{eqnarray}
It is easy to verify that both the classical action, $\cal S$, and the
inner product, are invariant under matrix Yang-Mills transformations.

\vskip 0.1in
Closure of the group of transformations that are the finite
$N$ manifestation of large $N$ continuum supersymmetry algebra
is a nontrivial result. However, as we will see below, with the
ordering prescription given earlier, the manipulations required 
to verify that $\cal S$ is supersymmetry invariant are well-defined.
Consider infinitesimal spinor parameters, $\eta_1$, $\eta_2$, each 
of which transforms as a $N$-vector of the
unitary group $SU(N)$. We must verify that the commutator
of two matrix supersymmetry transformations
with arbitrary infinitesimal spinor parameters
can always be expressed as the sum of
(i) an infinitesimal tangent space translation with parameter,
$\xi^{a}$$=$${\bar{\eta_1}} \Gamma^{a}
\eta_2$, (ii) an infinitesimal
local Lorentz transformation with parameter
${\rm L}_{bc}$$=$$\xi^{a} \omega_{abc}$, and (iii)
an infinitesimal local gauge transformation with gauge
parameter $\alpha^i$$=$$-g \xi^{a} A_{a}^i$ \cite{fs}.
Guided by the form of the locally supersymmetric type I action \cite{br},
we arrive at the following sequence of matrix transformations induced by
the infinitesimal matrix \lq\lq spinor" parameter, $\eta$, an
$N$-dimensional vector under $SU(N)$:
\begin{eqnarray}
\delta A^i_{\mu}  &&= \half {\bar{\eta}} \Gamma_{\mu} \chi^i
\cr
\delta \chi &&= - \half  ( \Gamma^{ab} F_{ab} ) \eta
+ \half \left ( {\bar{\eta}} \chi - {\bar{\chi}} \eta \right ) \lambda
- \half ({\bar{\chi}} \Gamma^a \eta ) \Gamma_a \lambda
\cr
\delta E^a_{\mu} &&= \half {\bar{\eta}} \Gamma^a \psi_{\mu}
\cr
\delta \psi_{\mu} &&= D_{\mu} \eta + \half \left (
{\bar{\eta}} \psi_{\mu} - {\bar{\psi}}_{\mu} \eta \right ) \lambda
 - \half  ( {\bar{\psi}}_{\mu} \Gamma^a \eta ) \Gamma_a \lambda
+ {{1}\over{g^2}} ({\bar{\chi}} \Gamma^{abc} \chi ) \Gamma_{abc} \Gamma_{\mu} \eta
\cr
\delta \Phi &&= {\bar{\eta}} \lambda
\cr
\delta \lambda &&=  -  \quart (\Gamma^a D_a \Phi  ) \eta +
\left ( H_{abc} - {\bar{\lambda}} \Gamma_{abc} \lambda
+ {{1}\over{g^2}}  {\rm tr} ( {\bar{\chi}} \Gamma_{abc} \chi ) \right )
 \Gamma^{abc} \eta
\quad .
\label{eq:susyt}
\end{eqnarray}
We emphasize that there is no ambiguity in the ordering of variables in 
the matrix supersymmetry transformation laws as given above.

\vskip 0.1in
In closing, we should note that, in principle, ${\cal S}$ belongs to a family of 
matrix actions, members of which can differ by $1/N$ corrections, thus yielding the 
same spacetime effective action in the infrared in accordance with the 
principle of Universality
Classes \cite{wilson,ktohwa}. Our procedure for determining $\cal S$ ensures that all
relevant interactions in the large $N$ continuum action that are 
required in order to match correctly with
the type I effective action with manifest Yang-Mills invariance, local supersymmetry, and
Lorentz invariance at the scale $\alpha^{\prime -1/2}$, are already present in the 
ultraviolet theory defined by ${\cal S}$. Thus, the sole source for both
nonperturbative, or quantum, corrections to the spacetime low energy effective action
of the type I string are the quantum corrections from the matrix path integral. 

\vskip 0.1in
We should emphasize that unlike the well-understood case of the hermitian one-matrix 
model where the angular variables decouple \cite{bipz}, the quantum corrections to
our matrix action will be sensitive to the off-diagonal entries of the $SU(N)$ matrix 
variables because of the proliferation of terms in the action coupling 
distinct species of matrices. However, unlike the case of a generic 
multi-matrix model, our classical action can be motivated largely by symmetry considerations 
alone. This important observation rests on the existence of the finite $N$ matrix symmetry
transformations demonstrated in this paper. We save further discussion of the matrix 
quantum effective action for future work.

\vskip 0.3in
\centerline{\bf ACKNOWLEDGMENTS}

\vskip 0.1in
\noindent
I am grateful to Hikaru Kawai for patient explanations of many 
aspects of planar reduction and the large $N$ limit of matrix theories. 
This work was completed in part during visits at Kyoto University, 
Cornell University, and at the Kavli Institute for Theoretical Physics, 
UCSB. It is funded in part by the NSF Career award NSF-PHY-9722394. 
 
\vskip 0.4in 
\noindent{\bf NOTE ADDED (JULY 2005):}  

\vskip 0.1in \noindent The basic framework given here
has been fleshed out in subsequent papers, notably hep-th/0408057 
[Nucl.\  Phys.\  {\bf B719} (2005) 188]. As noted in footnote 2 of this reference, the 
assignment of supersymmetric partners, bosonic and fermionic, to distinct $SU(N)$ 
representations,
namely, adjoint and vector, 
made in this earlier paper, was dropped by us in subsequent work. The reason 
is that, in such matrix models, supersymmetry does not
commute with the $SU(N)$ algebra, necessitating a rather 
nontrivial large $N$ limit: the correspondence to continuum target 
spacetime physics becomes obscure. This is also
the reason we invoke $SU(N)$, distinguishing the 
$N$ and ${\bar N}$ representations,
in this earlier work, while the adjoint representation of the
$U(N)$ flavor group is the only species of matrix variable 
mentioned in subsequent papers. Recall that either unitary or hermitian
matrices were acceptable in the $c$ $\le$ $1$ matrix models \cite{mat}; the distinction 
was found to be a 
moot point in the large $N$ continuum limit of smooth Riemann surfaces, a 
property that was 
called {\em matrix universality}. One should note that the discretization of 
unorientable worldsheets 
does require complex matrix models. In contrast, the matrix Lagrangians given here,
and in hep-th/0408057, 
 make no 
reference to discretized worldsheets: they match directly to the 
{\em target spacetime} formulation of superstring theory. But perhaps use can be made of this
broad class of generalized matrix models in other contexts. I would like to
thank Bernard de Wit for discussions of these important distinctions. The discussion
of electric-magnetic, and strong-weak coupling, dualities is
a bit confusing in this paper; the reader will wonder why the full spectrum
of Ramond-Ramond pform potentials is invoked at certain points in the
discussion, even though at other points we only refer
 to the unoriented type I-I$^{\prime}$ string theories in the large
$N$ continuum limit. Notice that we have implicitly included the heterotic string theories,
since they share the same perturbative string spectrum, and target space
Lagrangian, modulo duality transformations, in backgrounds with sixteen 
supercharges. And the generic type II pform potentials will enter into any description
of type IIA or type IIB backgrounds with sixteen supercharges.
Thus, the matrix Lagrangian must include all necessary pform 
potentials in order to incorporate these additional string backgrounds as 
suitable large $N$ limits. These aspects became
clearer to me in follow-up works: hep-th/0202138, 0205306, and 0210134.
Finally, note that a derivation of the matrix Lagrangian appearing in this paper, by 
the spacetime reduction technique, is given in hep-th/0408057.
A summary emphasizing electric-magnetic duality appears in hep-th/0507116.

\end{document}